\documentclass[twocolumn,aps,pre,amsmath,longbibliography,nofootinbib]{revtex4-2}
\usepackage{amsmath,amssymb}
\usepackage{graphicx}
\usepackage[colorlinks=true, bookmarks=true,
bookmarksnumbered=true, bookmarkstype=toc, linkcolor=magenta,
urlcolor=cyan, citecolor=cyan]{hyperref}

\begin{document}
\title{Instability of the ferromagnetic phase under random fields in an Ising spin glass with correlated disorder}
\author{Hidetoshi Nishimori}
\affiliation{Institute of Integrated Research, Institute of Science Tokyo, Nagatsuta-cho, Midori-ku, Yokohama 226-8503, Japan}
\affiliation{Graduate School of Information Sciences, Tohoku University, Sendai 980-8579,
Japan}
\affiliation{RIKEN Interdisciplinary Theoretical and Mathematical Sciences (iTHEMS),
Wako, Saitama 351-0198, Japan}
. \date{\today}
\begin{abstract} 
It is well established that the ferromagnetic phase remains stable under random magnetic fields in three and higher dimensions for the ferromagnetic Ising model and the Edwards-Anderson model of spin glasses without correlation in the disorder variables. In this study, we investigate an Ising spin glass with correlated disorder and demonstrate that the ferromagnetic phase becomes unstable under random fields in any dimension, provided that magnetic field chaos exists in the Edwards-Anderson model on the same lattice.  Additionally, we show that this instability can also be attributed to disorder (bond) chaos. We further argue that the model with correlated disorder remains in the ferromagnetic phase even in the presence of symmetry-breaking fields, as long as the Edwards-Anderson model on the same lattice exhibits a spin glass phase under a magnetic field.  These results underscore the profound impact of spatial correlations in the disorder.

\end{abstract}
\maketitle
\section{Introduction}
\label{sec:introduction}

Random fields applied to the ferromagnetic Ising model tend to destabilize the system, posing the intriguing question of whether the ferromagnetic phase can persist at low temperatures under such perturbations \cite{Young1997book}. This problem was first raised by Imry and Ma~\cite{Imry1975}, with subsequent rigorous proofs provided in Refs.~\cite{Imbrie1984,Bricmont1987,Bricmont1988,Aizenman1989,Aizenman1990,Ding2024,Ding2024a,Affonso2023}, which established the stability of the ferromagnetic phase in three and higher dimensions for the pure (non-random) Ising model. Similarly, it is widely believed that the ferromagnetic phase of the Edwards-Anderson model of spin glasses \cite{Edwards1975}, where disorder variables lack spatial correlations, remains stable in three and higher dimensions under the influence of random fields~\cite{Andresen2013,Andresen2017,Soares1994,Migliorini1998,Erichsen2021}\footnote{We use the terms `disorder' and `randomness' interchangeably. The reason is that `disorder' is used for consistency with our previous work~\cite{Nishimori2024}, while the term `random' is widely accepted for random fields.}. This suggests that the ferromagnetic phase of the Edwards-Anderson model shares essential characteristics with the pure ferromagnetic Ising model, at least with respect to its response to random fields. Moreover, experimental efforts to realize random field spin models have been actively pursued~\cite{Wu1993,Belanger1993,Wernsdorfer2005,Schechter2006,Tabei2006,Schechter2008,Andresen2014}.

Recent advances in understanding the effects of spatial correlations in disorder variables in spin glasses have uncovered unexpected features of the ferromagnetic phase in the Ising spin glass when a specific type of relatively strong correlation is introduced~\cite{Nishimori2024} in contrast to the case of weak correlations, which do not induce qualitative changes in the system properties~\cite{Hoyos2011,Bonzom2013,Cavaliere_2019,Munster2021,Nishimori2022}. For strongly correlated disorder of the type introduced in Ref.~\cite{Nishimori2024}, the magnetization distribution function exhibits support on a finite interval, provided that the spin glass phase of the Edwards-Anderson model on the same lattice has replica symmetry breaking~\cite{Parisi1980}. This implies that the magnetization is not self-averaging, deviating from the conventional expectation that the magnetization is self-averaging and its distribution function consists of only two delta functions at $\pm m_{\rm s}$ in the ferromagnetic phase, where $m_{\rm s}$ is the spontaneous magnetization.
Moreover, it has been demonstrated that the ferromagnetic phase is confined to a single line in the phase diagram of the correlated model, the Nishimori line (NL) \cite{Nishimori1980,Nishimori1981,Nishimori2001}, if temperature chaos, characterized by a drastic change in the spin state with slight temperature variations \cite{Bray1987,Banavar1987,Fisher1986,Fisher1988,Kondor1989,Ney-Nifle_1997,Ney-Nifle1998,Parisi2010,Mathieu2001,Bouchaud2001,Aspelmeier2002,Rizzo2003,Houdayer2004,Katzgraber2007b,Fernandez_2013,Wang2015,Billoire2018,Baity-Jesi2021}, exists in the Edwards-Anderson model on the same lattice.  This is probably the only example, in which the ferromagnetic phase on a single line in the phase diagram is surrounded by a non-ferromagnetic (spin glass) phase.

In this paper, we extend the findings of Ref.~\cite{Nishimori2024} by investigating the stability of the ferromagnetic phase on the NL under random fields. By augmenting the formulation of Ref.~\cite{Nishimori2024} to include random fields, we reach a striking conclusion: The ferromagnetic phase on the NL is unstable under symmetrically distributed random fields in any dimension, provided that magnetic field chaos~\cite{Parisi1983,Kondor1989,Ritort1994,Marinari2000,Billoire2003,Doussal2012,Chen2014,Aguilar2024}, which is a drastic change in the spin state with slight field variations, exists in the Edwards-Anderson model on the same lattice. Furthermore, we argue that this instability can also be interpreted as a manifestation of disorder (bond) chaos~\cite{Azcoiti_1995,Ney-Nifle1998,Sasaki2005,Krzakala_2005,Katzgraber2007b,Chen2013,Wang2016,Chen2018,Wang2018,Chatterjee2023}, which is a drastic change in the spin state with slight variations in the interactions.
We also find that the ferromagnetic phase persists even in the presence of symmetry-breaking fields, contrasting sharply with the behavior of a pure ferromagnet, where symmetry-breaking fields immediately suppress the ferromagnetic phase and replace it with the paramagnetic phase.

These results reveal the fundamentally different characteristics of the ferromagnetic phase in the correlated disorder model of the present type
compared to the Edwards-Anderson model without correlations in the disorder, suggesting that spatial correlations in the disorder profoundly alter the system properties.

This paper is organized as follows. In the next section, we formulate the problem and derive an identity relating the distribution functions of the magnetization and the replica overlap, which is then used to establish several non-trivial results. The final section is devoted to the conclusion.

\section{Ising spin glass in random fields with correlated disorder}
\subsection{The case without external fields}
To lay the groundwork for the subsequent discussion, we review the formulation of the problem for the case without fields examined in Ref.~\cite{Nishimori2024}. Readers already familiar with this reference may skip this subsection.

The analysis focuses on the dimensionless Hamiltonian,
\begin{align}
    H = -\beta \sum_{\langle ij \rangle} \tau_{ij}S_i S_j.
\end{align}
Here, $S_i(=\pm 1)$ is the Ising spin at site $i$ and $\tau_{ij}(=\pm 1)$ denotes the disordered interaction for the bond $\langle ij \rangle$ on an arbitrary lattice with an arbitrary range of interactions. The parameter $\beta$ corresponds to the coupling strength (inverse temperature).

The probability distribution function of the quenched disorder variables $\tau=\{\tau_{ij}\}$ is chosen to be
\begin{align}
    P(\tau) = \frac{1}{A} \frac{e^{\beta_p \sum_{\langle ij \rangle} \tau_{ij}}}{Z_{\tau}(\beta_p)},
    \label{eq:P_tau}
\end{align}
where $A$ is the normalization factor and $\beta_p$ is a control parameter. The denominator is the partition function
\begin{align}
    Z_{\tau}(\beta_p) = \sum_S e^{\beta_p \sum_{\langle ij \rangle} \tau_{ij}S_i S_j}.
\end{align}
The distribution Eq.~(\ref{eq:P_tau}) cannot be decomposed into the product of independent distributions,
\begin{align}
    P(\tau)\ne \prod_{\langle ij\rangle} p(\tau_{ij})
\end{align}
for some function $p(\cdot)$, unlike the standard Edwards-Anderson model
\begin{align}
    P_{\rm EA}(\tau)=\frac{e^{\beta_p\sum_{\langle ij\rangle}\tau_{ij}}}{(2\cosh \beta_p)^{N_{\rm B}}}=\prod_{\langle ij\rangle}\frac{e^{\beta_p\tau_{ij}}}{2\cosh \beta_p},
\end{align}
where $N_{\rm B}$ is the number of bonds,
and therefore $P(\tau)$ represents correlated disorder.

The two distributions coincide in the limit $\beta_p\to 0$,
\begin{align}
    \lim_{\beta_p\to 0}P(\tau)=\lim_{\beta_p\to 0}P_{\rm EA}(\tau)=\frac{1}{2^{N_{\rm B}}},
\end{align}
in which case $\tau_{ij}$ assumes $+1$ and $-1$ with the same probability. In the opposite limit $\beta_p\to \infty$, the Edwards-Anderson model $P_{\rm EA}(\tau)$ picks up only the purely ferromagnetic bonds $\tau_{ij}=1~(\forall \langle ij\rangle)$, whereas $P(\tau)$ behaves differently,
\begin{align}
    \lim_{\beta_p\to \infty}P(\tau)\ne\lim_{\beta_p\to \infty}P_{\rm EA}(\tau).
\end{align}
More precisely, the distribution $P(\tau)$ picks up not only the purely ferromagnetic disorder configuration $\tau_{ij}=1~(\forall \langle ij\rangle)$ but also those configurations with perfect ferromagnetic {\em spin} state $S_i=1~\forall i$ (or $S_i=-1~\forall i$) as one of the ground states. The reason is that if a disorder configuration whose ground state is not purely ferromagnetic survives in the limit $\beta_p\to\infty$, its ground-state energy $E_{\rm g}$ should be lower than the purely ferromagnetic energy $-\sum\tau_{ij}$ by $-\Delta E(<0)$,
\begin{align}
    E_{\rm g}=-\sum\tau_{ij}-\Delta E.
\end{align}
The probability of this disorder configuration in the limit $\beta_p\to\infty$ is, with $c$ being the ground state degeneracy,
\begin{align}
    P(\tau)\to \frac{1}{A}\,\frac{e^{\beta_p\sum\tau_{ij}}}{c\,e^{\beta_p(\sum\tau_{ij}+\Delta E)}}\to 0.
\end{align}
Consequently, these disorder configurations do not survive in this limit.

\begin{figure}[t]
\centering
\includegraphics[width=50mm]{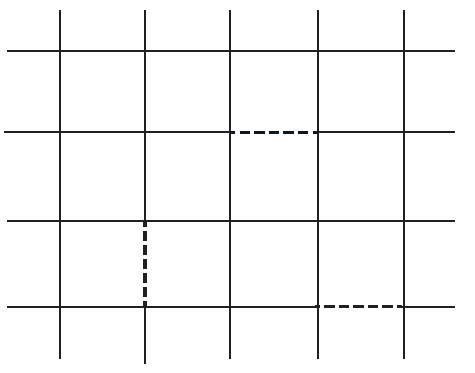}
\caption{One of the disorder configurations that survives in the limit $\beta_p\to\infty$ in the absence of external fields. Full lines represent ferromagnetic bonds $\tau_{ij}=1$, and dashed lines are for antiferromagnetic bonds $\tau_{ij}=-1$. The ground state for this disorder configuration has all spins aligned up (or all down).
}
\label{fig:bond_config}
\end{figure}

An example of a disorder configuration to survive in the limit $\beta_p\to \infty$ is shown in Fig.~\ref{fig:bond_config}.

\subsection{The problem with finite fields}
\label{subsec:definition}
We now generalize the problem to the case with random fields,
\begin{align}
    H = -\beta \sum_{\langle ij \rangle} \tau_{ij}S_i S_j - h \sum_{i=1}^{N} \mu_i S_i,
\end{align}
where $\mu_i(=\pm 1)$ is the disordered orientation of the magnetic field applied to site $i$, and $N$ is the number of sites. 
Generalizing Eq.~(\ref{eq:P_tau}), we introduce the probability distribution function as
\begin{align}
    P(\tau, \mu) = \frac{1}{A} \frac{e^{\beta_p \sum_{\langle ij \rangle} \tau_{ij} + h_p \sum_i \mu_i}}{Z_{\tau, \mu}(\beta_p, h_p)},
    \label{eq:P_tau-mu}
\end{align}
where $Z_{\tau, \mu}(\beta_p, h_p)$ is the partition function given by
\begin{align}
    Z_{\tau, \mu}(\beta_p, h_p) = \sum_S e^{\beta_p \sum_{\langle ij \rangle} \tau_{ij}S_i S_j + h_p \sum_i \mu_i S_i}.
\end{align}
It is seen that $Z_{\tau, \mu}(\beta_p, 0)$ is equal to $Z_\tau(\beta_p)$. The normalization factor $A$ can be evaluated by applying a gauge transformation $\tau_{ij} \to \tau_{ij}\sigma_i \sigma_j, \mu_i \to \mu_i \sigma_i~(\sigma_i=\pm 1)$ and summing the result over $\sigma = \{\sigma_i\}$, leading to
\begin{align}
    A &= \sum_{\tau, \mu} \frac{e^{\beta_p \sum_{\langle ij \rangle} \tau_{ij} + h_p \sum_i \mu_i}}{Z_{\tau, \mu}(\beta_p, h_p)} \nonumber \\
    &= \frac{1}{2^N} \sum_{\tau, \mu} \frac{\sum_\sigma e^{\beta_p \sum_{\langle ij \rangle} \tau_{ij}\sigma_i\sigma_j + h_p \sum_i \mu_i \sigma_i}}{Z_{\tau, \mu}(\beta_p, h_p)} \nonumber \\
    &= 2^{N_{\rm B}}.
    \label{eq:Ap}
\end{align}
Equation~(\ref{eq:P_tau-mu})  realizes weak correlations in the disorder variables for small $\beta_p$ and $h_p$ and strong correlations for large $\beta_p$ and $h_p$.  We are interested in the case with large $\beta_p$, where the ferromagnetic phase exists.

The discussions in the next section extend straightforwardly to the Gaussian distribution function by replacing summations over $\tau$ and $\mu$ with integrals over Gaussian variables $J$ and $\mu$,
\begin{align}
    &\frac{1}{2^{N_{\rm B} + N}}\sum_{\tau, \mu} f(\tau, \mu) 
    \nonumber\\
    &\longrightarrow \int_{-\infty}^\infty f(J, \mu) \prod_{\langle ij \rangle} \frac{e^{-\frac{{J_{ij}}^2}{2}}}{\sqrt{2\pi}} dJ_{ij} \prod_i \frac{e^{-\frac{\mu_i^2}{2}}}{\sqrt{2\pi}} d\mu_i.
    \label{eq:gaussian}
\end{align}
\subsection{Main identity}
\label{subsec:main_identity}
One of the quantities of interest is the probability distribution function of the magnetization
\begin{align}
    P_1(x|\beta,h,\beta_p,h_p)=\Big[\big\langle \delta\big(x-\frac{1}{N}\sum_iS_i\big)\big\rangle_{\beta,h}\Big]_{\beta_p,h_p},
\end{align}
where $\langle \cdots\rangle_{\beta,h}$ denotes the thermal average and $[\cdots ]_{\beta_p,h_p}$ is for the configurational average by the probability $P(\tau,\mu)$. More explicitly,
\begin{align}
    &P_1(x|\beta,h,\beta_p,h_p)
    =\frac{1}{A}\sum_{\tau,\mu}\frac{e^{\beta_p \sum \tau_{ij}+h_p\sum\mu_i}}{Z_{\tau,\mu}(\beta_p,h_p)}
   \nonumber\\
    &\cdot\frac{\sum_S \delta\big(x-\frac{1}{N}\sum_iS_i\big)e^{\beta \sum \tau_{ij}S_iS_j+h\sum\mu_iS_i}}{Z_{\tau,\mu}(\beta,h)}\label{eq:P1}.
\end{align}
We introduce an additional probability distribution for the overlap of two replicas with different parameters $\beta_1, h_1$ and $\beta_2,h_2$, averaged over the common disorder configuration $P(\tau,\mu)$,
\begin{widetext}
\begin{align}
&P_2(x|\beta_1,h_1,\beta_2,h_2)=\Big[\big\langle \delta\big(x-\frac{1}{N}\sum_iS_i^{(1)}S_i^{(2)}\big)\big\rangle_{\beta_1,h_1,\beta_2,h_2}\Big]_{\beta_p,h_p}
\nonumber\\
    &=\frac{1}{A}\sum_{\tau,\mu}\frac{e^{\beta_p \sum \tau_{ij}+h_p\sum_i\mu_i}}{Z_{\tau,\mu}(\beta_p,h_p)}
    \,\frac{\sum_{S^{(1,2)}} \delta\big(x-\frac{1}{N}\sum_iS_i^{(1)}S_i^{(2)}\big)e^{\beta_1 \sum \tau_{ij}S_i^{(1)}S_j^{(1)}+h_1\sum\mu_iS_i^{(1)}}e^{\beta_2 \sum \tau_{ij}S_i^{(2)}S_j^{(2)}+h_2\sum\mu_iS_i^{(2)}}}{Z_{\tau,\mu}(\beta_1,h_1)Z_{\tau,\mu}(\beta_2,h_2)}
\end{align}
\end{widetext}
If we apply the gauge transformation $\tau_{ij} \to \tau_{ij} \sigma_i \sigma_j, \mu_i \to \mu_i \sigma_i, S_i^{(1)} \to S_i^{(1)} \sigma_i, S_i^{(2)} \to S_i^{(2)} \sigma_i$ to the second line of the above equation and sum the result over $\sigma = \{\sigma_i\}$, we obtain
\begin{widetext}
\begin{align}
    P_2(x|\beta_1,h_1,\beta_2,h_2)=\frac{1}{2^{N}A}\sum_{\tau,\mu}
    \frac{\sum_{S^{(1,2)}} \delta\big(x-\frac{1}{N}\sum_iS_i^{(1)}S_i^{(2)}\big)e^{\beta_1 \sum \tau_{ij}S_i^{(1)}S_j^{(1)}+h_1\sum\mu_iS_i^{(1)}}e^{\beta_2 \sum \tau_{ij}S_i^{(2)}S_j^{(2)}+h_2\sum\mu_iS_i^{(2)}}}{Z_{\tau,\mu}(\beta_1,h_1)Z_{\tau,\mu}(\beta_2,h_2)}.
    \label{eq:P2}
\end{align}
\end{widetext}
Equation (\ref{eq:P2}) demonstrates that $P_2(x|\beta_1,h_1,\beta_2,h_2)$ is independent of $\beta_p$ and $h_p$, and thus we have omitted these variables from the argument of $P_2$.

The notation $P_{1,2}(x|\cdots)$ indicates that $P_1$ and $P_2$ are functions of $x$, conditioned on the specified hyperparameters $(\beta,h,\beta_p,h_p)$ or $(\beta_1,h_1,\beta_2,h_2)$. Equation (\ref{eq:P2}) shows that $P_2(x|\beta_1,h_1,\beta_2,h_2)$ represents the replica overlap of two replicated systems, each characterized by inverse temperatures and fields $(\beta_1, h_1)$ and $(\beta_2, h_2)$, respectively, for the Edwards-Anderson model with independent and symmetric distributions for $\tau$ and $\mu$. 

The randomness in the fields $\mu$ in Eq.~(\ref{eq:P2}) can be eliminated through the gauge transformation $\tau_{ij} \to \tau_{ij} \mu_i \mu_j, S_i^{(1)} \to S_i^{(1)} \mu_i, S_i^{(2)} \to S_i^{(2)} \mu_i$. Consequently, the replica overlap of the Edwards-Anderson model with a symmetric distribution of random fields is equivalent to the replica overlap under uniform fields.

The main identity of this paper is the following relation,
\begin{align}
    P_1(x|\beta,h,\beta_p,h_p) &= P_2(x|\beta,h,\beta_p,h_p) \nonumber \\
        &= P_2(x|\beta_p,h_p,\beta,h).
    \label{eq:p1p2}
\end{align}
The first equality is readily derived by applying the gauge transformation to Eq.~(\ref{eq:P1}). The second equality is a straightforward consequence of Eq.~(\ref{eq:P2}). Equation (\ref{eq:p1p2}) generalizes the identity derived and analyzed in Ref.~\cite{Nishimori2024} to the present case with random fields. It demonstrates that the distribution function of the magnetization for the model with correlated disorder is equal to the replica overlap of the Edwards-Anderson model with a symmetric distribution of independent disorder.

Despite the simplicity of its derivation, the above identity should be regarded as a highly non-trivial relation, since it directly equates the property of ferromagnetism in the correlated model of the present type with that of spin glasses in the uncorrelated Edwards-Anderson model. We will see its profound consequences in the following.

\subsection{Physical consequences of the identity}
\begin{figure}[t]
\centering
\includegraphics[width=65mm]{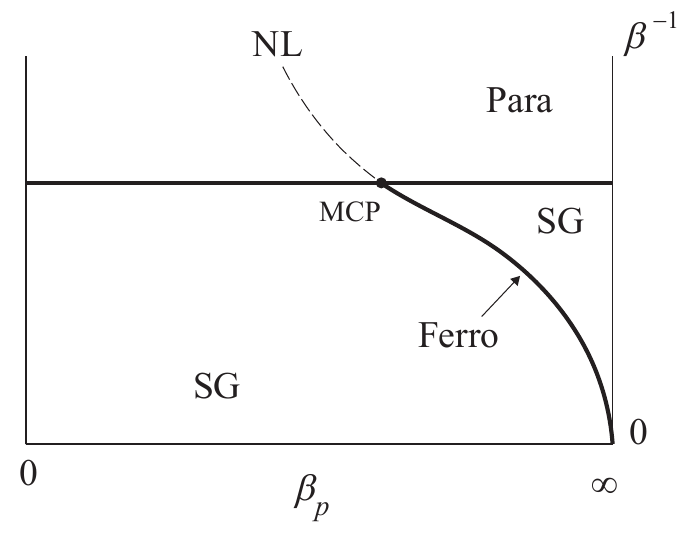}
\caption{In the absence of external fields $h=h_p=0$, the ferromagnetic phase exists only on the NL when the Edwards-Anderson model ($\beta_p=0$) has temperature chaos. MCP is the multicritical point.
}
\label{fig:phase_diagram3}
\end{figure}

It has been demonstrated in Ref.~\cite{Nishimori2024} using Eq.~(\ref{eq:p1p2}) for the case $(h_p = h = 0)$ that the model with correlated disorder but no random fields exhibits anomalous behavior on and near the NL (defined by $\beta_p = \beta$) in the ferromagnetic phase, provided the Edwards-Anderson model with symmetric disorder on the same lattice exhibits replica symmetry breaking \cite{Parisi1980} and/or temperature chaos \cite{Bray1987,Banavar1987,Fisher1986,Fisher1988,Kondor1989,Ney-Nifle_1997,Ney-Nifle1998,Parisi2010,Mathieu2001,Bouchaud2001,Aspelmeier2002,Rizzo2003,Houdayer2004,Katzgraber2007b,Fernandez_2013,Wang2015,Billoire2018,Baity-Jesi2021,Dahlberg2024}.
In particular, the distribution function of the magnetization $P_1(x|\beta,0,\beta,0)$ has support on a finite interval as in the distribution function of the replica overlap of the Parisi type \cite{Parisi1980}, and the ferromagnetic phase is restricted to the NL  defined by $\beta=\beta_p$ when temperature chaos exists as illustrated in Fig.~\ref{fig:phase_diagram3}.

We discuss in the present paper the case with finite fields and show that the potential presence of magnetic field chaos (hereafter referred to  as field chaos for simplicity)  in the Edwards-Anderson model gives rise to a different type of anomaly in the model with correlated disorder of the present type.

Before delving into detailed discussions, it is useful to confirm that the model without random fields exhibits a ferromagnetic phase on the NL (i.e., $m(\beta,0,\beta,0)\ne 0$) at intermediate and large $\beta$ if the Edwards-Anderson model possesses a spin glass phase $q(\beta,0,\beta,0)\ne 0$ at the same temperature. Here, the magnetization and the spin glass order parameter are defined as
\begin{align}
    m(\beta,h,\beta_p,h_p)&=\int_{-1}^1
    dx\, xP_1(x|\beta,h,\beta_p,h_p)\label{Eq:m_def}\\
    q(\beta,h,\beta_p,h_p)&=\int_{-1}^1 dx \,xP_2(x|\beta,h,\beta_p,h_p)\label{eq:q_def}.
\end{align}
The magnetization is finite $m(\beta,0,\beta,0)\ne 0$ if $q(\beta,0,\beta,0)\ne 0$, because the above definitions, together with  Eq.~(\ref{eq:p1p2}), lead to
\begin{align}
    m(\beta,0,\beta,0)=q(\beta,0,\beta,0)\ne 0.
    \label{eq:mq0}
\end{align}
Although a more general relation  holds, 
\begin{align}
    m(\beta,h,\beta_p,h_p)=q(\beta,h,\beta_p,h_p),
\end{align}
$q(\beta,h,\beta_p,h_p)$ has the meaning of the spin glass order parameter only when the two replicas share the same parameters ($\beta=\beta_p$ and $h=h_p$), which is the NL condition. Note that fixed boundary conditions are imposed here to avoid the trivial vanishing of $m(\beta,0,\beta,0)$ and $q(\beta,0,\beta,0)$  by the $\mathbb{Z}_2$ symmetry, but this is not necessarily the case when $h\ne 0$ and also when we consider the distribution functions, not the average values as in Eqs.~(\ref{Eq:m_def}) and (\ref{eq:q_def}).

\subsubsection{Unstable ferromagnetic phase under random fields}
\label{subsec:field-chaos}
We now examine the implications of Eq.~(\ref{eq:p1p2}) under the conditions $\beta = \beta_p$, $h > 0$, and $h_p = 0$,
\begin{align}
P_1(x|\beta, h, \beta, 0) = P_2(x|\beta, h, \beta, 0). \label{eq:p1p20}
\end{align}
Here, the left-hand side represents the distribution function of the magnetization for the model with correlated disorder on the NL $(\beta = \beta_p)$ under symmetrically distributed $(h_p = 0)$ random fields $(h>0)$. The right-hand side is the distribution function of the overlap between two replicas of the Edwards-Anderson model, one subjected to random fields ($h > 0$) and the other without (the final argument being zero, $h_p = 0$).

Field chaos is a distinctive phenomenon proposed to occur in spin glasses \cite{Kondor1989, Ritort1994, Billoire2003, Doussal2012, Chen2014, Aguilar2024}, in which even the slightest change in the strength of an external field induces a drastic reorganization of the spin state. More precisely, if field chaos is present, the spin configurations of two replicas, one subjected to a finite external field $h$ and the other in the absence of a field, become entirely uncorrelated. Consequently, the overlap between the spin configurations of the two replicas, represented by the right-hand side of Eq.~(\ref{eq:p1p20}), collapses to a single delta peak at $x=0$ for any $h > 0$ in the thermodynamic limit,
\begin{align}
P_2(x|\beta, h, \beta, 0) = \delta(x). \label{eq:p2_delta}
\end{align}
Therefore, Eq.~(\ref{eq:p1p20}) implies
\begin{align}
P_1(x|\beta, h, \beta, 0) = \delta(x), \label{eq:p2_delta2}
\end{align}
and thus
\begin{align}
m(\beta, h, \beta, 0) = 0. \label{eq:m=0}
\end{align}
This equation proves that, regardless of the spatial dimensionality and any other conditions on the lattice structure, the ferromagnetic phase, which existed on the NL in the model with correlated disorder in the absence of random fields as described in Eq.~(\ref{eq:mq0}), disappears upon the introduction of unbiased ($h_p=0$) random fields, if field chaos exists in the Edwards-Anderson model on the same lattice as defined in Eq.~(\ref{eq:p2_delta}). This instability of the ferromagnetic phase is a fundamentally different property compared to the pure Ising model and the Edwards-Anderson model, where the ferromagnetic phase remains stable in three and higher dimensions under random fields \cite{Imry1975,Imbrie1984, Bricmont1987, Bricmont1988,Aizenman1989,Aizenman1990,Ding2024,Ding2024a,Affonso2023, Andresen2013, Andresen2017, Soares1994, Migliorini1998, Erichsen2021}.

It is worth noting that $h$ in Eq.~(\ref{eq:m=0}) can be arbitrarily small as long as it stays finite. If we wish to take the limit $h \to 0$, it must be carried out after the thermodynamic limit. Otherwise, different behaviors may emerge \cite{Aguilar2024}.

When both $h$ and $h_p$ are finite but they are not equal to each other $h\ne h_p$, there can still exist field chaos. In this case, the overlap of two replicas has a single delta peak away from the origin \cite{Billoire2003},
\begin{align}
    P_1(x|\beta,h,\beta,h_p)=\delta(x-q_m(\beta,h,h_p))~~(h\ne h_p),\label{eq:p1-trivial}
\end{align}
where $q_m$ is some function of $\beta$, $h$, and $h_p$. 
The shift of the delta peak from $x = 0$ stems from the fact that the finite fields $h$ and $h_p$ impose a preferential alignment of the spin configurations, quantified by $q_m$. Beyond this trivial finite overlap, however, the two spin configurations remain completely different.

Notice that the above single delta distribution function is characteristic of the paramagnetic phase with a single trivial state.

\subsubsection{Disorder chaos}
We next present an alternative perspective to the  analysis in the previous subsection. 

In the absence of random fields ($h=h_p=0$), the distribution function of the magnetization under the NL condition ($\beta=\beta_p$) is given by
\begin{align}
   & P_1(x|\beta,0,\beta,0)\nonumber\\
    &=\frac{1}{A}\sum_{\tau,\mu}\frac{e^{\beta \sum \tau_{ij}}}{Z_{\tau}(\beta)}
   \,
    \frac{\sum_S \delta\big(x-\frac{1}{N}\sum_iS_i\big)e^{\beta \sum \tau_{ij}S_iS_j}}{Z_{\tau}(\beta)}\nonumber\\
    &=\frac{2^N}{A}\sum_{\tau}\frac{e^{\beta \sum \tau_{ij}}}{Z_{\tau}(\beta)}
   \,
    \frac{\sum_S \delta\big(x-\frac{1}{N}\sum_iS_i\big)e^{\beta \sum \tau_{ij}S_iS_j}}{Z_{\tau}(\beta)},
    \label{eq:U_dc0}
\end{align}
where the summation over $\mu$ has been taken in going to the last line.
As shown below, the introduction of symmetric random fields ($h>0, h_p=0$) modifies the distribution function of the interactions $\tau$, leading to
\begin{align}
 P_1(x|\beta,h,\beta,0)&=\frac{1}{A}\sum_{\tau}\frac{e^{\beta \sum \tau_{ij}}}{Z_{\tau}(\beta)}\, U(\tau|\beta,h)\nonumber\\
 &
   \cdot \frac{\sum_S \delta\big(x-\frac{1}{N}\sum_iS_i\big)e^{\beta \sum \tau_{ij}S_iS_j}}{Z_{\tau}(\beta)},
    \label{eq:U_dc}
\end{align}
for some function $U(\tau|\beta,h)$ with the property $U(\tau|\beta,0)=1$  but $U(\tau|\beta,h) \ne 1$ for $h \ne 0$ . Comparing Eqs.~(\ref{eq:U_dc0}) and (\ref{eq:U_dc}), we see that the introduction of a finite $h$ alters the distribution function of bond variables $\tau$. 
Furthermore, according to Eqs.~(\ref{eq:p1p20}) to (\ref{eq:p2_delta2}) in the previous subsection, the spin configurations of two replicas, one with finite $h>0$ and the other without field, are completely different if field chaos exists,
\begin{align}
    P_2(x|\beta,h,\beta,0)(=P_1(x|\beta,h,\beta,0))=\delta(x).
\end{align}
This implies that modifying the bond variable distribution from Eq.~(\ref{eq:U_dc0}) to Eq.~(\ref{eq:U_dc}) induces a drastic change in the spin state. 
Such sensitivity to perturbations in the bond disorder distribution precisely corresponds to the definition of disorder (bond) chaos \cite{Azcoiti_1995,Ney-Nifle1998,Sasaki2005,Krzakala_2005,Katzgraber2007b,Wang2016,Chen2018,Wang2018,Chatterjee2023}.

The proof of Eq.~(\ref{eq:U_dc}) is straightforward. According to Eq.~(\ref{eq:p1p2}), the arguments $(\beta,h)$ and $(\beta_p,h_p)$ in $P_1$ can be exchanged,
\begin{align}
   & P_1(\beta,h,\beta,0)=P_1(x|\beta,0,\beta,h)\nonumber\\
    &=\frac{1}{A}\sum_{\tau,\mu}\frac{e^{\beta \sum \tau_{ij}+h\sum\mu_i}}{Z_{\tau,\mu}(\beta,h)}
   \,
    \frac{\sum_S \delta\big(x-\frac{1}{N}\sum_iS_i\big)e^{\beta \sum \tau_{ij}S_iS_j}}{Z_{\tau}(\beta)}.\label{eq:h0}
\end{align}
Here, the summation over $\mu$ can be carried out independently of the thermal average over the $S$ variables,
\begin{align}
    \sum_{\mu}\frac{e^{\beta \sum \tau_{ij}+h\sum\mu_i}}{Z_{\tau,\mu}(\beta,h)}\equiv  e^{\beta \sum \tau_{ij}} U(\tau|\beta,h), \label{eq:mu-sum}
\end{align}
which defines the function $U(\tau|\beta,h)$.
In the limit of small $h$, the right-hand side reduces to
\begin{align}
    e^{\beta \sum \tau_{ij}}\lim_{h\to 0}U(\tau|\beta,h)=\frac{2^Ne^{\beta \sum \tau_{ij}}}{Z_{\tau}(\beta)}\label{eq:h02}
\end{align}
as is apparent from the definition. This ends the proof.

\subsubsection{Distribution function for the ferromagnetic ordering}
The third observation pertains to the distribution function of the ferromagnetic ordering, with the NL condition applied to both the inverse temperature and the field ($\beta_p=\beta$ and $h_p=h$),
\begin{align}
    P_1(\beta,h,\beta,h)=P_2(\beta,h,\beta,h).
    \label{eq:p1p2h}
\end{align}
The right-hand side represents the overlap between two replicas of the Edwards-Anderson model under the same values of applied fields. If a spin-glass phase with replica symmetry breaking exists in the Edwards-Anderson model under fields on a given lattice, the right-hand side will have support on a finite interval as long as the system remains in the spin-glass phase, that is, when $\beta$ is sufficiently large and $h$ is sufficiently small. Consequently, the left-hand side, which describes the distribution function of the magnetization for the model with disorder correlations, also has support on a finite interval.

It is generally believed that the magnetization distribution function, which does not involve multiple replicas, features at most two delta peaks. However, the present result challenges this understanding by providing an exception. This finding generalizes the case without random fields as presented in Ref.~\cite{Nishimori2024}, where the same conclusion was reached for the case $h=0$ that the distribution function of the magnetization of the present model has the same form as the distribution of the spin glass order parameter of the Edwards.

It is interesting that the present model with correlated disorder stays in the ferromagnetic phase under finite random fields if the NL condition is imposed both on the inverse temperature $\beta_p=\beta$ and the field $h_p=h$\footnote{We call it the ferromagnetic phase based on the fact $m(\beta,h,\beta,h)\ne 0$ under the non-trivial (non-single-delta) distribution of the magnetization $P_1(x|\beta,h,\beta,h)$.}. This is in contrast to the case of the previous subsections with $\beta=\beta_p$ but $h_p\ne h$, in which case the system is in the paramagnetic phase, see Eq.~(\ref{eq:p1-trivial}).

\begin{figure}[t]
\centering
\includegraphics[width=42mm]{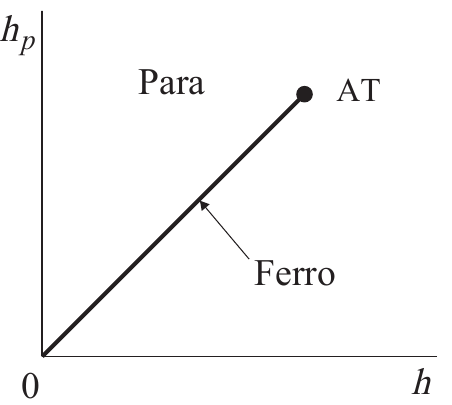}
\caption{Phase diagram of the model with disorder correlation with $\beta(=\beta_p)$ fixed to a sufficiently large value.  It is assumed that replica symmetry breaking and field chaos exist in the Edwards-Anderson model on the same lattice. The ferromagnetic phase exists only along the diagonal up to the point marked AT corresponding to the Almeida-Thouless line. The rest of the phase diagram is occupied by the paramagnetic phase.
}
\label{fig:phase_diagram}
\end{figure}

If the finite-dimensional Edwards-Anderson model has a similar property to the mean-field Sherrington-Kirkpatrick model \cite{Sherrington1975}, the stability condition of the spin glass phase under fields breaks down at higher temperatures (smaller $\beta$) and/or larger fields corresponding to the Almeida-Thouless line \cite{Almeida_1978}. Then, the right-hand side of Eq.~(\ref{eq:p1p2h}) becomes a single delta function, so does the left-hand side, implying that the present model with correlated disorder leaves the ferromagnetic phase and enters the paramagnetic phase, see Fig.~\ref{fig:phase_diagram}. This existence of the ferromagnetic phase in the present model for $h=h_p\ne 0$ is highly non-trivial because the model has symmetry-breaking fields to align the spins into the up direction ($h_p>0$). In the pure ferromagnetic Ising model, any small amount of a symmetry-breaking field will immediately drive the system away from the ferromagnetic phase, which is characterized by the distribution
\begin{align}
P(x|\beta, h=0)=\frac{1}{2}\{\delta(x-m_{\rm s}(\beta))+\delta(x+m_{\rm s}(\beta))\},
\end{align}
where $m_s$ is the spontaneous magnetization, into the paramagnetic phase with a single delta 
\begin{align}
    P(x|\beta,h)=\delta(x-m(\beta,h)).
\end{align}
Here $P(x|\beta,h)$ is the distribution function of the magnetization for the pure Ising model,
\begin{align}
    P(x|\beta,h)=\frac{\sum_S \delta(x-\frac{1}{N}\sum_i S_i)\,e^{\beta \sum S_iS_j+h\sum_i S_i}}{\sum_S e^{\beta \sum S_iS_j+h\sum_i S_i}}.
\end{align}

Figure \ref{fig:phase_diagram} summarizes the results derived in this subsection and Sec.~\ref{subsec:field-chaos} by illustrating the phase diagram of the present model in the $h$-$h_p$ plane with $\beta(=\beta_p)$ fixed to a sufficiently large value. It is assumed that the spin glass phase in random fields and field chaos both exist in the Edwards-Anderson model on the same lattice. The ferromagnetic phase with a non-trivial distribution function of the magnetization exists along the diagonal $h=h_p$ up to a point corresponding to the Almeida-Thouless line marked AT. The rest of the phase diagram ($h\ne h_p$) is occupied by the paramagnetic phase with the distribution function of the magnetization having a single delta peak. The existence of the ferromagnetic phase only on a line as a result of chaos is similar to the case of the temperature-probability phase diagram analyzed in the previous paper \cite{Nishimori2024}.

\subsection{More generic formulation}
It may be useful to present a more generic formulation at the end of this main section. Let us consider the probability distribution function defined by
\begin{align}
    &P_{\rm g}(x|\beta,\epsilon)=\nonumber\\
    &\frac{1}{A}\sum_{\tau,\mu}\frac{e^{\beta \sum \tau_{ij}}}{Z_{\tau}(\beta)}
   \,
    \frac{\sum_S \delta\big(x-\frac{1}{N}\sum_iS_i\big)e^{\beta \sum \tau_{ij}S_iS_j+\epsilon H_{\rm pert}(\tau,\mu)}}{Z_{\tau,\mu}(\beta,\epsilon)}.\label{eq:P-epsilon}
\end{align}
Here, $H_{\rm pert}(\tau,\mu)$ is the perturbation Hamiltonian with gauge invariance and $\epsilon$ is the small parameter. Examples include the random fields that we already discussed,
\begin{align}
    H_{\rm pert}=\sum_i \mu_i S_i,
\end{align}
and the original Hamiltonian,
\begin{align} 
    H_{\rm pert}=\sum \tau_{ij} S_i S_j.
\end{align}
The latter amounts to a small change in the inverse temperature $\beta\to \beta+\epsilon$ in the exponent of Eq.~(\ref{eq:P-epsilon}), which may lead to temperature chaos. A more general combination of $\{\tau_{ij}S_i S_j\}$ such as
\begin{align}
    H_{\rm pert}=\sum \xi_{ij}\tau_{ij}S_iS_j,
\end{align}
where each $\xi_{ij}$ is randomly fixed to an arbitrary value, may lead to disorder (bond) chaos. Another interesting case is the transverse field, in which case the summation over $S$ should be replaced by the trace in the Hilbert space,
\begin{align}
    &m(\beta,\epsilon)=\nonumber\\
    &\frac{1}{A}\sum_{\tau,\mu}\frac{e^{\beta \sum \tau_{ij}}}{Z_{\tau}(\beta)}
   \,
    \frac{
    {\rm Tr}\,\big(\frac{1}{N}\sum_i \sigma_i^z\big)e^{\beta \sum \tau_{ij}\sigma_i^z \sigma_j^z+\epsilon \sum_i \sigma_i^x}}{Z_{\tau}(\beta,\epsilon)},\label{eq:P-tfim}
\end{align}
where $\sigma_i^z$ and $\sigma_i^x$ are the $z$ and $x$ component of the Pauli matrix at site $i$, respectively\footnote{The gauge transformation for the Pauli matrix is $\sigma_i^x\to\sigma_i^x, \sigma_i^y\to \sigma_i \sigma_i^y, \sigma_i^z\to\sigma_i\sigma_i^z$, where $\sigma_i(=\pm 1)$ is the classical gauge variable.}. Note that a distribution function is difficult to define in this case because $\sum_i \sigma_i^z$ is an operator.

It is straightforward to show that $P_{\rm g}(x|\beta, \epsilon)$ (or $m(\beta,\epsilon)$ for the transverse-field Ising model) is equal to the distribution function of the overlap between two replicas (or the average overlap for the transverse-field Ising model), one with the perturbation $\epsilon H_{\rm pert}$ and the other without it, in the Edwards-Anderson model. If chaos is induced by $\epsilon H_{\rm pert}$ in the Edwards-Anderson model, the ferromagnetic phase on the NL in the model with correlated disorder is unstable under the introduction of the perturbation.

\section{Conclusion}
In the previous paper \cite{Nishimori2024}, we studied the effects of a specific type of correlation in the disorder variables for the Ising spin glass without random fields. It was shown that the ferromagnetic phase on the Nishimori line (NL) exhibits unusual properties, including support on a finite interval in the magnetization distribution function of the type of the Parisi distribution function of the spin glass order parameter as well as the confinement of the ferromagnetic phase strictly to the NL. These findings relied on the assumptions of the existence of the spin glass phase with replica symmetry breaking and temperature chaos in the Edwards-Anderson model on the same lattice.

In this paper, we have extended the theoretical framework to incorporate random fields, providing new insights into their influence on the system with correlated disorder of the same type as studied before. Specifically, we have proven that the ferromagnetic phase on the NL becomes unstable in the presence of symmetrically distributed random fields in any dimension, assuming the existence of field chaos in the Edwards-Anderson model on the same lattice. This result sharply contrasts with the behavior of the pure ferromagnetic Ising model and the Edwards-Anderson model, where the ferromagnetic phase is known to remain stable under random fields in three and higher dimensions as long as they are not too strong \cite{Imry1975,Imbrie1984,Bricmont1987,Bricmont1988,Aizenman1989,Aizenman1990,Ding2024,Ding2024a,Affonso2023,Andresen2013,Andresen2017,Soares1994,Migliorini1998,Erichsen2021}. Moreover, it has been shown that the instability of the ferromagnetic phase in the present model can also be seen as a consequence of the phenomenon of disorder (bond) chaos. In consideration of the likelihood of the existence of disorder (bond) chaos in the three-dimensional Edwards-Anderson model \cite{Katzgraber2007b}, the instability of the ferromagnetic phase on the NL of the present model under random fields is a realistic possibility in three dimensions. 

Additionally, the magnetization distribution function on the NL exhibits an anomalous feature with its support on a finite interval if the Edwards-Anderson model has replica symmetry breaking in finite fields. This fact is related to the property of the present model that it stays in the ferromagnetic phase even under symmetry-breaking external fields, in contrast to the pure ferromagnetic model where any small amount of symmetry-breaking field destroys the ferromagnetic phase. Conversely, if we stick to the conventional understanding that the ferromagnetic phase is replaced by the paramagnetic phase immediately after the introduction of symmetry-breaking fields, then the Edwards-Anderson model does not have a spin glass phase  at finite fields with replica symmetry breaking of the Parisi type.

We have also presented a generic formulation of the problem, which describes various types of chaos in a unified way.

The analyses in this paper make no assumptions about spatial dimensionality or the range of interactions and are therefore applicable to the mean-field Sherrington-Kirkpatrick model as well. However, it is important to note that in the present theory, the thermodynamic limit is generally taken after most of the computations, with the notable exception of the zero-field limit in Sec. \ref{subsec:field-chaos}. In contrast, mean-field computations require the thermodynamic limit to be taken as the first step of the analyses to take advantage of the saddle-point evaluation. Careful consideration is needed to determine whether or not this exchange of limits leads to restrictions on the applicability of the present results to the Sherrington-Kirkpatrick model.

The findings provided in this paper reveal that the ferromagnetic phase on the NL of the present model with correlated disorder behaves fundamentally differently compared to the pure ferromagnetic Ising model and the Edwards-Anderson model without correlation in the disorder. Further investigations including numerical studies are desirable to fully elucidate the properties of the ferromagnetic phase in the present model to deepen our understanding of the role of disorder correlations in spin glasses. 
Additionally, exploring other forms of spatial correlations in disorder variables would be valuable for determining the extent to which the properties observed in this model are general features of disorder correlation or specific to the particular structure considered here.

\acknowledgments
The author thanks Manaka Okuyama for helpful comments.

%

\end{document}